\documentclass[12pt]{article}
\usepackage[top=1in, bottom=1in, left=1in, right=1in]{geometry}
\usepackage{graphicx}
\usepackage{amsmath,amssymb}
\usepackage{color}
\usepackage{hyperref}
\usepackage{slashed}
\usepackage{ulem}

\newcommand{\den}{16+(3n_q-16)c_\beta^2}

\definecolor{darkblue}{rgb}{0,0,0.9}
\hypersetup{
	linktocpage,
	colorlinks,
	citecolor=darkblue,
	filecolor=darkblue,
	linkcolor=darkblue,
	urlcolor=darkblue,
	pdftitle={Colorphilic Spin-2 Resonances at the LHC},
	pdfauthor={R. Sekhar Chivukula, Dennis Foren, and Elizabeth H. Simmons}
}

\title{\bf Colorphilic Spin-2 Resonances in the LHC Dijet Channel}
\author{R. Sekhar Chivukula, Dennis Foren,  and Elizabeth H. Simmons}
\date{}

\begin{document}
\maketitle
\thispagestyle{empty}
\begin{center}
 {\it Department of Physics and Astronomy\\
 Michigan State University,\\
 East Lansing U.S.A.}
\end{center}

\section*{\hfil {Abstract} \hfil} 
Experiments at the LHC may yet discover a dijet resonance indicative of Beyond the Standard Model (BSM) physics. In this case, the question becomes: what BSM theories are consistent with the unexpected resonance? One possibility would be a spin-2 object called the ``colorphilic graviton"--a spin-2 color-singlet particle which couples exclusively to the quark and gluon stress-energy tensors. We assess the possibility of this state's discovery in the dijet channel as an s-channel resonance, and report the regions of parameter space where colorphilic gravitons have not yet been excluded by LHC-13 data but still may be discovered in the dijet channel at LHC-14 for integrated luminosities of 0.3, 1, and 3 ab$^{-1}$. We then delineate which of those regions remain accessible to future collider searches, once one accounts for applicability of the narrow-width approximation, mass resolution of the detector, and self-consistency according to tree-level partial-wave unitarity. We discover that--despite the strong constraints unitarity imposes on collider searches--the colorphilic graviton remains potentially discoverable in the LHC dijet channel. A means of investigation would be to apply the color discriminant variable, a dimensionless combination of quantities (production cross-section, total decay width, and invariant mass) that can be quickly measured after the discovery of a dijet resonance. Previous publications have demonstrated the color discriminant variable's utility when applied to theories containing vector bosons (colorons, $Z^\prime$), excited quarks, and diquarks. We extend this analysis to the case of the colorphilic graviton by applying the color discriminant variable to the appropriate region of parameter space. We conclude that resolvable, discoverable dijet resonances consistent with colorphilic gravitons span a narrower range of masses than those consistent with leptophobic $Z^\prime$ models, and can be distinguished from those originating from coloron, excited quark, and diquark models.

\newpage

\setcounter{page}{1}
\section{Introduction}
New heavy particles with sizable couplings to quarks and gluons also have relatively large production cross-sections at the Large Hadron Collider (LHC). Such particles could appear as resonances in the dijet channel above the otherwise rapidly-falling QCD background. In this vein, CMS and ATLAS are continually searching the dijet channel for evidence of new physics. Because a resonance has yet to be discovered, they report 95\% CL exclusion limits on various benchmark models. As more data is acquired and higher collision energies explored, the exclusion limits strengthen, eliminating larger classes of models. Our work utilizes CMS exclusion limits from an analysis of $\sim36$ fb$^{-1}$ of LHC-13 dijet channel data \cite{CMS2017}. These limits are comparable to ATLAS exclusion limits based on $\sim37$ fb$^{-1}$ of LHC-13 dijet channel data \cite{ATLAS2017}.

Discovery of a new dijet resonance $R$ would indicate a deviation from the Standard Model (SM) of particle physics, and signal the presence of physics beyond the Standard Model (BSM). In the event of a discovery, experiments will immediately measure the dijet cross-section of the resonance ($\sigma_{Rjj}$), its mass ($m_R$), and its total decay width ($\Gamma_R$) if possible. We would subsequently wonder what new physics is consistent with these quantities or combinations of these quantities. In principle, a new dijet resonance could have one of many different color and spin structures. In this work we will focus on spin-2 resonances.

Experimental searches typically choose Randall-Sundrum (RS) gravitons as their spin-2 benchmark model \cite{Randall:1999ee}. RS gravitons appear as Kaluza-Klein excitations of extra-dimensional gravitational theories. Because RS gravitons couple to the full Standard Model stress-energy tensor, their dijet and leptonic couplings are of comparable size, and RS gravitons are more likely to be discovered through leptonic channels than dijet channels at hadron colliders. Leptonic channels are cleaner than dijet channels, largely because they do not have to compete against the large QCD background.

However, the literature also includes other BSM spin-2 models, such as the various spin-2 resonances described in Refs \cite{Artoisenet:2013puc}-\cite{Han:2015cty}. A generic graviton may couple to any number of stress-energy tensors. We use the term ``colorphilic graviton" (labeled $X_2$) to describe a phenomenological massive spin-2 object that couples exclusively to the quark and gluon stress-energy tensors with strengths proportional to parameters $\kappa_q$ and $\kappa_g$ respectively. By construction, a colorphilic graviton is likely to be first observed in the dijet channel as an $s$-channel resonance. This parallels other particles to which the color discriminant variable has been applied. Explicit specification of a UV completion for the colorphilic graviton models we consider is beyond the scope of this article.

Phenomenological spin-2 particles warrant caution, because their interactions generically violate unitarity at high enough collision energies \cite{Artoisenet:2013puc}. The colorphilic graviton $X_2$ couples to $q\overline{q}$ and $gg$ states through dimension-5 operators, necessitating couplings $\kappa_q$ and $\kappa_g$ with units of inverse energy. For small partonic center-of-momentum collision energy $\sqrt{\hat{s}}$ between pairs of incoming partons, this effective theory is valid order-by-order in powers of $\kappa\sqrt{\hat{s}}$. At sufficiently high partonic scattering energy, $\kappa\sqrt{\hat{s}}$ grows large enough to destroy perturbativity of the effective theory, as manifested in unitarity violation. 

Our work here on production and decay of colorphilic gravitons focuses on two-into-two scattering processes. To ensure the validity of our cross-section calculations, we use a tree-level partial-wave unitarity analysis to enforce unitarity. Similar analyses have been applied to other spin-2 objects in the literature \cite{Kim:2001rc}-\cite{Han:2004wt}. We demonstrate how this further constrains the parameter space accessible to collider searches and find that there exist regions of parameter space within which a colorphilic graviton is discoverable in the 14 TeV LHC dijet channel.

Suppose a new dijet resonance $R$ is discovered within the range of parameter space where a colorphilic graviton is accessible. How could we immediately distinguish whether $R$ is a spin-2 state? The dimensionless color discriminant variable provides a means of quickly diagnosing what classes of BSM models might describe a newly discovered resonance:
\begin{align}
D_{col,R} \equiv \dfrac{m_R^3\sigma_{Rjj}}{\Gamma_R}
\label{eq:dcoldefn}
\end{align}
This is constructed as to be independent of coupling strength when applied to $s$-channel resonances with narrow widths. The color discriminant variable is so-named because it is proportional to the color and spin structure of the resonance. Color-singlet objects and color-octet objects tend to be well separated in the color discriminant variable vs resonance mass plane \cite{Atre:2013mja}. 

Previous analyses have demonstrated the color discriminant variable's ability to discriminate between color-singlet, color-octet, and excited quark dijet resonances \cite{Atre:2013mja}-\cite{Chivukula:2014pma}. The method has also been broadened to models with flavor-dependent couplings \cite{Chivukula:2014npa}-\cite{Simmons:2015gna} and applied to scalar diquarks \cite{Chivukula:2015zma}. The present article applies the color discriminant variable to a spin-2 particle.

The remainder of the paper is organized in the following manner. Section 2 describes the colorphilic graviton, $X_2$. Section 3 details the constraints (including unitarity) that are relevant to identifying an LHC-discoverable dijet resonance originating from a colorphilic graviton and plots the surviving parameter space. Section 4 reviews the uncertainties relevant to a color discriminant variable analysis, plots the color discriminant variable (including uncertainties) vs the mass of $X_2$, and then compares the $X_2$ analysis to an equivalent $Z^\prime$ analysis. Section 5 summarizes our conclusions.

\section{Colorphilic Graviton Models}
This section presents the colorphilic graviton, its dijet cross-section, total decay width, and color discriminant variable.

\subsection{Lagrangian and Parameterization}
A colorphilic graviton is defined to be a massive spin-2 object $X_2$ that is a SM singlet and couples exclusively to particles charged under the SM $SU(3)_C$. We write the Lagrangian density interaction terms describing how $X_2$ couples to the Standard Model as follows:
\begin{align}\label{eq:Lag}
\mathcal{L}_{int} =  -\dfrac{\kappa_g}{2}X_2^{\mu\nu}T^g_{\mu\nu}-\sum_{q_i}\dfrac{\kappa_{q_i}}{2}X_2^{\mu\nu}T^{q_i}_{\mu\nu}
\end{align}
where $X_2^{\mu\nu}$ labels the $X_2$ field, $T^q_{\mu\nu}$ and $T^g_{\mu\nu}$ are the quark and gluon stress-energy tensors,
\begin{align}
T^g_{\mu\nu} &= \tfrac{1}{4}\eta_{\mu\nu}F^{\rho\sigma}F_{\rho\sigma}- {F_\mu}^\rho F_{\nu\rho} -\tfrac{1}{\xi}\eta_{\mu\nu}  \left[\partial^\rho \partial^\sigma A_{\sigma} A_{\rho} + \dfrac{1}{2}(\partial^\rho A_\rho)^2\right]\nonumber\\
&\hspace{20 pt}+\tfrac{1}{\xi}(\partial_\mu\partial^\rho A_\rho A_\nu + \partial_\nu \partial^\rho A_\rho A_\mu)\\
T^{q_i}_{\mu\nu} &= -\eta_{\mu\nu}\overline{q}_i\left(i\gamma^\rho D_\rho  - m_\psi\right)q_i+\tfrac{i}{2}\overline{q}_i(\gamma_\mu D_\nu +\gamma_\nu D_\mu)q_i+\tfrac{i}{2}\eta_{\mu\nu}\partial^\rho(\overline{q}_i\gamma_\rho q_i) \nonumber\\
&\hspace{20 pt}-\tfrac{i}{4}\partial_\mu(\overline{q}_i\gamma_\nu q_i)-\tfrac{i}{4}(\overline{q}_i\gamma_\mu q_i)
\end{align}
and $A$ and $q_i$ are the gluon and quark fields respectively (color indices have been suppressed) \cite{Han:1998sg}. We normalize the spin-2 polarization tensors $\epsilon^s_{\mu\nu}$ according to the published version of Ref \cite{Han:1998sg}, such that,
\begin{align}
\epsilon^{s,\mu\nu}\epsilon^{s^\prime *}_{\mu\nu} = \delta^{ss^\prime}
\end{align}
where $s\in\{-2,-1,0,1,2\}$ is a helicity index and $\mu,\nu$ are Lorentz indices. Eq. (\ref{eq:Lag}) implies tree-level couplings for the following interactions: $q_i\overline{q}_iX_2$, $ggX_2$, $q_i\overline{q}_igX_2$, $q_i\overline{q}_iV X_2$, $gggX_2$, and $ggggX_2$, where $V$ denotes the electroweak bosons $\gamma, W^\pm, Z$.

For the present analysis, we reduce the number of free variables by assuming $X_2$ has flavor universal couplings to quarks ($\kappa_{q_i}=\kappa_q$ for every quark $q_i$). This assumption allows us to simplify the present analysis and avoid flavor constraints. Consequently, there are only two couplings in the theory: the quark stress-energy coupling $\kappa_q$ and the gluon stress-energy coupling $\kappa_g$. These couplings have units of inverse energy. By using the mass of $X_2$ (denoted $m_{X_2}$) as an energy scale, we exchange the dimensionful couplings $\kappa_q,\kappa_g\in[0,+\infty)$ TeV for the dimensionless parameters $\alpha\in[0,+\infty)$ and $\beta\in[0,\tfrac{\pi}{2}]$, which are defined according to,
\begin{align}
\kappa_q \equiv \dfrac{\alpha}{m_{X_2}}\cos\beta \hspace{75 pt}\kappa_g \equiv \dfrac{\alpha}{m_{X_2}}\sin\beta
\end{align}
The cases where $X_2$ couples exclusively to $T^q_{\mu\nu}$, where $X_2$ couples equally to $T^q_{\mu\nu}$ and $T^g_{\mu\nu}$, and where $X_2$ couples exclusively to $T^g_{\mu\nu}$ correspond to $\beta=0$, $\beta=\tfrac{\pi}{4}$, and $\beta =\tfrac{\pi}{2}$ respectively.  Let $c_\beta$ and $s_\beta$ denote $\cos\beta$ and $\sin\beta$ respectively. Eq. (\ref{eq:Lag}) yields the following tree-level decay widths:
\begin{align}
\Gamma_{X_2\rightarrow gg} = \dfrac{\alpha^2 m_{X_2}}{40 \pi}s^2_\beta\hspace{50 pt}\Gamma_{X_2\rightarrow q_i \overline{q}_i} = \dfrac{3\alpha^2 m_{X_2}}{640\pi}c^2_\beta \left(1+\dfrac{8m_{q_i}^2}{3m_{X_2}^2}\right)\left[1-\dfrac{4m_{q_i}^2}{m^2_{X_2}}\right]
\end{align}
The model described by Eq. (\ref{eq:Lag}) additionally permits three-body decays $X_2\rightarrow q_i\overline{q}_ig$, $X_2\rightarrow q_i\overline{q}_iV$, and $X_2\rightarrow ggg$ where $V$ denotes the electroweak bosons ($\gamma$, $W^\pm$, $Z$), as well as the four-body decay $X_2\rightarrow gggg$. After accounting for infrared divergences in processes involving multiple massless bosons, these decay channels are numerically negligible relative to the two-body decay widths due to additional coupling and phase space factors. Analysis of these additional decay channels lies outside of our tree-level dijet channel analysis, and we ignore them for the remainder of this article.\footnote{With sufficient luminosity, the three- and four-body decay channels might provide a means to distinguish the colorphilic graviton from other models, such as the leptophobic $Z^\prime$.}

The total decay width is well described as the sum of the $gg$ and $q\overline{q}$ channels:
\begin{align}\label{eq:totwidth}
\Gamma_{X_2} &= \dfrac{\alpha^2 m_{X_2}}{640 \pi}\left(16s^2_\beta + 3 n_q c^2_\beta \right)
\end{align}
where $n_q$ acts as an effective number of massless quarks,
\begin{align}
n_q \equiv \sum_{q_i}\left(1+\dfrac{8m_{q_i}^2}{3m_{X_2}^2}\right)\left[1-\dfrac{4m_{q_i}^2}{m^2_{X_2}}\right]
\end{align}
In the limit where $m_{X_2}\gg m_{q_i}$ for every quark flavor, we find $n_q\rightarrow 6$. Only the top quark provides significant deviations from the massless case when $m_{X_2}>500\text{ GeV}$, which is the mass range relevant to our analysis. As a result, $n_q$ is well approximated by the following expression:
\begin{align}
n_q = 5 + \left( 1+\dfrac{8m_{t}^2}{3m_{X_2}^2}\right)\left[1-\dfrac{4m_{t}^2}{m^2_{X_2}}\right]
\end{align}
where $m_t$ is the mass of the top quark. As $m_{X_2}$ increases, $n_q$ quickly approaches $6$. For example, $n_q(1\text{ TeV}) = 5.89$ and $n_q(2\text{ TeV}) = 5.97$. We may write the relevant branching ratios as,
\begin{align}\label{eq:partonBr}
Br(X_2\rightarrow gg) &= \dfrac{16s^2_\beta}{\den}\\
Br(X_2\rightarrow q_i\overline{q}_i) &= \dfrac{3 c^2_\beta}{\den}\cdot \begin{cases}
1\hspace{51 pt}q_i=u,d,s,c,b\\
n_q-5\hspace{25 pt}q_i=t
\end{cases}
\end{align}
after some simplification.

\subsection{Colorphilic Gravitons in the Dijet Channel}
We are interested in decays of $X_2$ to pairs of light jets $j$, which we define as a QCD jet originating from a partonic gluon or one of the five lightest quarks. At tree-level, the decay width and branching ratio to light dijets equal,
\begin{align}\label{eq:jjwidth}
\Gamma_{X_2\rightarrow jj} = \dfrac{\alpha^2 m_{X_2}}{640\pi}(16- c^2_\beta)\hspace{50 pt}Br(X_2\rightarrow jj) = \dfrac{16-c_\beta^2}{\den}
\end{align}
Searches for narrow dijet resonances by CMS and ATLAS have greatest sensitivity to resonances with width-over-mass ratios below $\sim 0.15$ \cite{NWA}, which is also the range of values wherein the narrow width approximation (NWA) applies \cite{Harris:2011bh}. The dijet cross-section of a resonance $R$ with mass $m_R$ equals, in this approximation,
\begin{align}\label{eq:XS}
\sigma_{Rjj} = 16\pi^2\dfrac{\Gamma_R}{m_R}Br(R\rightarrow jj)\left\{\sum_{i,k}(1+\delta_{ik})\mathcal{N}_{ik}Br(R\rightarrow ik)\left[\dfrac{1}{s}\dfrac{dL^{ik}}{d\tau}\right]_{s\tau = m_R^2}\right\}
\end{align}
The indices $i,k$ in Eq. (\ref{eq:XS}) label partons in each of the incoming protons. At tree level, $X_2$ couples to gluons, quarks, and antiquarks, so that in principle $i,k\in\{g$, $u$, $d$, $s$, $c$, $b$, $t$, $\overline{u}$, $\overline{d}$, $\overline{s}$, $\overline{c}$, $\overline{b}$, $\overline{t}\}$. Because the proton has negligible top quark content and $X_2$ couples diagonally to quark-antiquark pairs, we restrict $\overline{k}=i\in\{g$, $u$, $d$, $s$, $c$, $b\}$. The contribution of a partonic combination $i,k$ to proton-proton collisions is described by its corresponding parton luminosity function,
\begin{align}
\left[\dfrac{d\mathcal{L}^{ik}}{d\tau}\right]\equiv \dfrac{1}{1+\delta_{ik}}\int_\tau^1 \dfrac{dx}{x}\left[f_i(x,\mu_F^2)f_k\left(\dfrac{\tau}{x},\mu_F^2\right)+f_k(x,\mu_F^2)f_i\left(\dfrac{\tau}{x},\mu_F^2\right)\right]
\end{align}
where $\mu_F$ is the factorization scale and $f_i(x,\mu_F^2)$ is the parton distribution function for parton $i$ \cite{Harris:2011bh}. We set the factorization scale to the resonance mass such that $\mu_F = m_{X_2}$ and evaluate the parton luminosity functions at $\tau = s/m_{X_2}^2$, where $s$ is the proton-proton center-of-momentum energy squared. We use the parton distribution functions from the CTEQ6L1 PDF set \cite{Pumplin:2002vw} for our calculations, and take care when extracting data from other sources to use each source's choice of PDF set.

The factor $\mathcal{N}_{ik}$ in Eq. (\ref{eq:XS}) counts color and spin degrees of freedom for the partonic combination $i,k$ relative to the resonance $R$:
\begin{align}
\mathcal{N}_{ik} \equiv \dfrac{N_{S_R}}{N_{S_i}N_{S_k}}\cdot \dfrac{C_R}{C_iC_k}
\end{align}
$N_S$ and $C$ are a given particle's number of spin and color states respectively. $(N_{S},C)$ equals $(5,1)$, $(2,3)$, and $(2,8)$ for the colorphilic graviton, quarks and antiquarks, and gluons respectively, yielding,
\begin{align}\label{eq:NN}
\mathcal{N}_{q_i\overline{q}_i} = \dfrac{5}{36}\hspace{75 pt}\mathcal{N}_{gg} = \dfrac{5}{256}
\end{align}
These considerations allow us to simplify Eq. (\ref{eq:XS}) to,
\begin{align}
\sigma_{X_2jj}&= 16\pi^2 \dfrac{\Gamma_{X_2\rightarrow jj}}{m_{X_2}^3}\left\{2\mathcal{N}_{gg}Br(X_2\rightarrow gg)\left[\tau \dfrac{dL^{gg}}{d\tau}\right]_{s\tau = m_{X_2}^2} +\right. \nonumber\\
&\hspace{150 pt}\mathcal{N}_{q_i\overline{q}_i}Br(X_2\rightarrow u\overline{u})\left.\sum_{q_k = u}^b\left[\tau \dfrac{dL^{q_k \overline{q}_k}}{d\tau}\right]_{s\tau = m_{X_2}^2}\right\}\label{eq:midXS}
\end{align}
Substituting Eqs. (\ref{eq:partonBr}-\ref{eq:jjwidth}) and Eq. (\ref{eq:NN}) into Eq. (\ref{eq:midXS}) yields an explicit expression for the dijet cross-section in terms of $m_{X_2}$, $\alpha$, $\beta$, and $s$.
\begin{align}\label{eq:NWAXS}
\sigma_{X_2jj}=\dfrac{(16-c_\beta^2)\pi  \alpha^2}{192[\den]m_{X_2}^2}\left\{3s_\beta^2\left[\tau \dfrac{dL^{gg}}{d\tau}\right]\ + 2c_\beta^2\sum_{q_i = u}^b\left[ \tau\dfrac{dL^{q_i \overline{q}_i}}{d\tau}\right]\right\}_{s\tau = m_{X_2}^2}
\end{align}
The color discriminant variable $D_{col,X_2}$ is, therefore,
\begin{align}
D_{col,X_2} &\equiv \dfrac{m_{X_2}^3\sigma_{X_2jj}}{\Gamma_{X_2}} \\
&= \dfrac{10\pi^2(16-c_\beta^2)}{3[\den]^2}\left\{3s_\beta^2\left[\tau \dfrac{dL^{gg}}{d\tau}\right]\ + 2c_\beta^2\sum_{q_i = u}^b\left[ \tau\dfrac{dL^{q_i \overline{q}_i}}{d\tau}\right]\right\}_{s\tau = m_{X_2}^2}\label{eq:Dcol}
\end{align}
Eqs. (\ref{eq:NWAXS}-\ref{eq:Dcol}) are valid for the process $pp\rightarrow X_2\rightarrow jj$, where $j$ is a light jet originating from a partonic $g$, $u$, $d$, $s$, $c$, or $b$.

\section{Parameter Space}
In this section, we describe the region of parameter space where the colorphilic graviton might be detected by the LHC with $0.3$, $1$, and $3$ ab$^{-1}$ of LHC-14 integrated luminosity for $\beta=0$, $\tfrac{\pi}{4}$, and $\tfrac{\pi}{2}$, and describe the subregion where a color discriminant variable analysis of such a discovery would apply.

Note the $\beta=0$ case corresponds to $X_2$ coupling exclusively to the quark stress-energy tensor, whereas the $\beta=\tfrac{\pi}{2}$ case corresponds to $X_2$ coupling exclusively to the gluon stress-energy tensor. When plotting the parameter space of the colorphilic graviton, we fix $\beta$ and plot $\alpha$ vs $m_{x_2}$. As discussed in Section 3.3, enforcing tree-level unitarity of a $X_2$ model establishes a scale $\Lambda^{\text{EFT}}_{max}$ up to which our effective field theory respects tree-level unitarity. This scale can be used as an additional parameter to constrain an effective theory. Therefore, we concern ourselves with four parameters: the colorphilic graviton mass $m_{X_2}$; a unitless coupling strength $\alpha$; an angle $\beta$ measuring the relative coupling strength of $X_2$ to quarks vs gluons; and an upper limit $\Lambda_{max}^{\text{EFT}}$ on partonic center-of-momentum energies for which the theory respects tree-level partial wave unitarity.

We eliminate regions of parameter space that are already experimentally 95\% CL excluded and only consider regions of parameter space where $5\sigma$ dijet resonance discovery may eventually be observed at LHC-14. The theory must respect unitarity to be self-consistent, leading us to consider unitarity constraints obtained from tree-level partial-wave amplitudes. Narrow dijet resonance searches by CMS and ATLAS have limited sensitivity for dijet resonances with $\Gamma_R/m_R\gtrsim 0.15$, and so we eliminate these regions of parameter space as well \cite{NWA}. 

Because the color discriminant variable analysis is appropriate in regions of parameter space where the width $\Gamma_{X_2}$ of $X_2$ is wide enough to be measured by the detector, we also consider how $\Gamma_{X_2}$ compares to the detector's mass resolution $M_{res}$.

\subsection{Excluded Region}
As mentioned in the introduction, CMS and ATLAS are searching for resonances in the dijet channel, and, in the absence of a resonance--establishing 95\% CL exclusion limits on dijet resonances \cite{CMS2017}-\cite{ATLAS2017}. Both experiments report these limits as upper bounds on acceptance $A_R$ times dijet cross-section $\sigma_{Rjj}$ as a function of resonance mass $m_R$. They also plot $A_R\times \sigma_{Rjj}$ for several benchmark models. The exclusion limits presented by ATLAS and CMS are comparable; we choose to utilize the CMS exclusion limits, which are obtained from 36 fb$^{-1}$ of LHC-13 dijet channel data.

The acceptance $A_{2}[\beta]$ of the colorphilic graviton is calculated for each value of $\beta$ in {\sc{MadGraph 5}} \cite{MadGraph} according to the cuts described in Ref \cite{CMS2017}. We then demand,
\begin{align}
A_2[\beta]\cdot \sigma_{X_2jj} \leq (A\times\sigma_{jj})_{95\%\text{CL}}
\end{align}
The region of parameter space that fails to satisfy this constraint, {\it i.e.} where a colorphilic graviton is excluded by current LHC data at 95\% CL, is located in the upper left area of each plot in Fig. \ref{ParameterSpace}, bounded by a thick black curve, and colored with a translucent dark red. All other points of each plot have not been excluded by the limits of Ref \cite{CMS2017}.

\subsection{5$\sigma$ Discovery Reach}
We also establish regions of parameter space where a dijet resonance might someday be discovered. The CMS experiment has published how sensitive their detector is towards $5\sigma$ discoveries in the dijet channel for $\mathcal{L}_{int} = 100\text{ pb}^{-1}$, $1\text{ fb}^{-1}$, and $10\text{ fb}^{-1}$ \cite{Gumus:2006mxa}. This is reported as an upper limit on dijet cross-section times acceptance. We calculate the relevant spin-2 acceptance via {\sc{MadGraph 5}} \cite{MadGraph} for each value of $\beta$ according to the cuts described in Ref \cite{Gumus:2006mxa}.

We assume the systematic uncertainties scale proportionally to the square root of integrated luminosity $\sqrt{\mathcal{L}_{int}}$ to extend the $10\text{ fb}^{-1}$ discovery prospects of Ref \cite{Gumus:2006mxa} to $\mathcal{L}_{int}=0.3$, $1$, and $3\text{ ab}^{-1}$. Any areas of parameter space that require more than $3\text{ ab}^{-1}$ worth of LHC-14 data according to this scaling are designated as inaccessible and we exclude that region of parameter space. Naive scaling provides a qualitative idea of LHC-14 discovery prospects; however, we expect the regions we plot ultimately to be conservative because this scaling ignores any improved experimental sensitivities in the high luminosity LHC-14 dijet channel.

In each pane of Fig. \ref{ParameterSpace}, the boundary of every $5\sigma$ discovery region is denoted with a black curve, while the regions themselves are denoted in white. Curves corresponding to larger values of $\mathcal{L}_{int}$ are further rightward, with $\mathcal{L}_{int}=0.3$, $1$ and $3\text{ ab}^{-1}$ appearing from left to right respectively. The region above and to the left of a given $\mathcal{L}_{int}$ curve corresponds to the region of parameter space accessible with $\mathcal{L}_{int}$ worth of LHC-14 dijet channel data. The gray region to the bottom-right of each plot is the previously-described inaccessible region of parameter space, which requires more than $3\text{ ab}^{-1}$ worth of LHC-14 dijet channel data to discover a colorphilic graviton.

\subsection{Unitarity Constraints}
Because $X_2$ couples to quarks and gluons via dimension-5 operators, its couplings are dimensionful and the colorphilic graviton generically violates unitarity once the energy scales of the process exceed some energy scale $\Lambda_{max}^{\text{EFT}}$ \cite{Artoisenet:2013puc}. This is indicative of the breakdown of the effective field theory (EFT) approximation implicit in our analysis. Once unitarity is violated, the effective field theory becomes invalid and the method of analysis must be changed. If an EFT $\mathcal{L}_{\text{EFT}}$ arises as an approximation of a more fundamental perturbative theory $\mathcal{L}_{fund}$, then its breakdown must be circumvented with additional new physics effects relevant to $\mathcal{L}_{fund}$. These new physics effects become relevant at some energy scale $\Lambda_{NP}$. In order that $\mathcal{L}_{fund}$ respect unitarity at energies higher than $\mathcal{L}_{\text{EFT}}$, the new physics must enter before the EFT's breakdown, such that $\Lambda_{NP} <\Lambda^{\text{EFT}}_{max}$. We may demand a model containing a colorphilic graviton respect tree-level unitarity up to some value of $\Lambda_{max}^{\text{EFT}}$ and in doing so we ensure that our analysis including only a colorphilic graviton is self-consistent at energies relevant to collider searches, and the underlying new physics could be at higher energies than probed by the LHC.

Unitarity demands certain relationships between matrix elements, including constraints on $2\rightarrow 2$ elastic scattering amplitudes mediated by $s$-channel spin-2 particles such as the colorphilic graviton. In particular, each eigenvalue $a_2$ of the partial-wave amplitude matrix corresponding to this process must satisfy $|\mathfrak{R}[a_2]|\leq 1/2$. We calculate the tree-level $2_a\rightarrow X_2 \rightarrow 2_b$ amplitudes where $2_a,2_b\in \{q_i\overline{q}_i,gg\}$ and resulting constraint in Appendix B. One consequence of this constraint is that $\Gamma_{X_2}/m_{X_2}$ must be smaller than $\tfrac{1}{8}$.

We fix $\Lambda_{max}^{\text{EFT}}$ and seek regions of parameter space where a colorphilic graviton model satisfies Eq. (\ref{eq:unicon}) for values of $\sqrt{\hat{s}}$ up to at least $\Lambda_{max}^{\text{EFT}}$. Combining Eq. (\ref{eq:smax}) and Eq. (\ref{eq:simpcombo}) with Eq. (\ref{eq:unicon}), for each value of $\Lambda_{max}^{\text{EFT}}$ we demand $m_{X_2}$, $\alpha$, and $\beta$ satisfy:
\begin{align}
\Lambda^{\text{EFT}}_{max} < m_{X_2}\left[\dfrac{80\pi}{(8+c_{\beta}^2)\alpha^2}\left(1+\sqrt{1-(8+c_{\beta}^2)\dfrac{\alpha^2}{40\pi}}\right)\right]^{1/2}
\end{align}
In each pane of Fig. \ref{ParameterSpace}, we plot the boundary of this constraint for $\Lambda_{max}^{\text{EFT}}=10$, $33$, and $100$ TeV. Each boundary is linear at low resonance mass $m_{X_2}$ and denoted by a red line, with shallower slopes corresponding to larger values of  $\Lambda_{max}^{\text{EFT}}$. The region above a given $\Lambda_{max}^{\text{EFT}}$ boundary would encounter unitarity violation for some value of $\sqrt{\hat{s}}$ below $\Lambda_{max}^{\text{EFT}}$. The lowest value of $\Lambda^{\text{EFT}}_{max}$ we might consider setting is $\Lambda_{max}^{\text{EFT}}=m_{X_2}$, so as to ensure consistency of the theory up to production of the $X_2$ particle. Beyond this consideration, the choice of $\Lambda_{max}^{\text{EFT}}$ is arbitrary. Based on the ranges of masses admitted by the other constraints in the analysis, the largest $m_{X_2}$ discoverable with $3$ ab$^{-1}$ of LHC-14 data and consistent with $\Gamma_{X_2}/m_{X_2}\leq \tfrac{1}{8}$ is $m_{X_2}\sim 5\text{ TeV}$. Because we assume each colorphilic graviton model is a low-energy approximation of a more fundamental perturbative theory, new physics must become relevant at some scale below $\Lambda_{max}^{\text{EFT}}$. We choose EFT consistency up to $\Lambda_{max}^{\text{EFT}}=10$ TeV for the remainder of the analysis, as to allow room for new physics above $m_{X_2}$, and we exclude from our analysis points in the translucent red region above the $\Lambda_{max}^{\text{EFT}}=10$ TeV boundary line.

\subsection{Narrow Width Approximation}
As described in Section 2.2, we utilize the narrow width approximation, which provides a good approximation when $\Gamma_{X_2}/m_{X_2} \leq 0.15$ \cite{NWA}-\cite{Harris:2011bh}. This inequality implies an upper limit on $\alpha$:
\begin{align}\label{eq:NWAcon}
\alpha \leq \sqrt{\dfrac{(0.15)640\pi}{16+(3n_q-16)c_\beta^2}}
\end{align}
However, as discovered in Appendix B, unitarity demands $\Gamma_{X_2}/m_{X_2}\leq 1/8$, which is a stronger constraint. Therefore, enforcement of tree-level partial wave unitarity ensures colorphilic graviton models remain in the regime of the narrow width approximation. The simple form of Eq. \ref{eq:NWAcon} motivates plotting $X_2$ parameter space as $(m_{X_2},\alpha)$ instead of $(m_{X_2},\kappa)$.

The boundary at which $\Gamma_{X_2}/m_{X_2} = 0.15$ is marked by a horizontal black line near the top of each pane of Fig. \ref{ParameterSpace}. The grayed region directly above this line corresponds to $\Gamma_{X_2}/m_{X_2}>0.15$ and is excluded from our analysis.

\subsection{Mass Resolution}
The dijet mass resolution $M_{res}$ of a detector determines whether experiments can measure the total decay width $\Gamma_R$ of a particular dijet resonance. Specifically, the resonance width is measurable if it exceeds the mass resolution; else, the resonance width is irresolvable. Because the color discriminant variable is explicitly constructed from the resonance width of a dijet resonance, it is only applicable to identifying the nature of the resonance when $\Gamma_R\geq M_{res}$. Dijet resonances for which $\Gamma_R < M_{res}$ may still be detected, but would require a different kind of identification analysis. 

\begin{figure}[p]
\begin{center}
\includegraphics[scale=0.60]{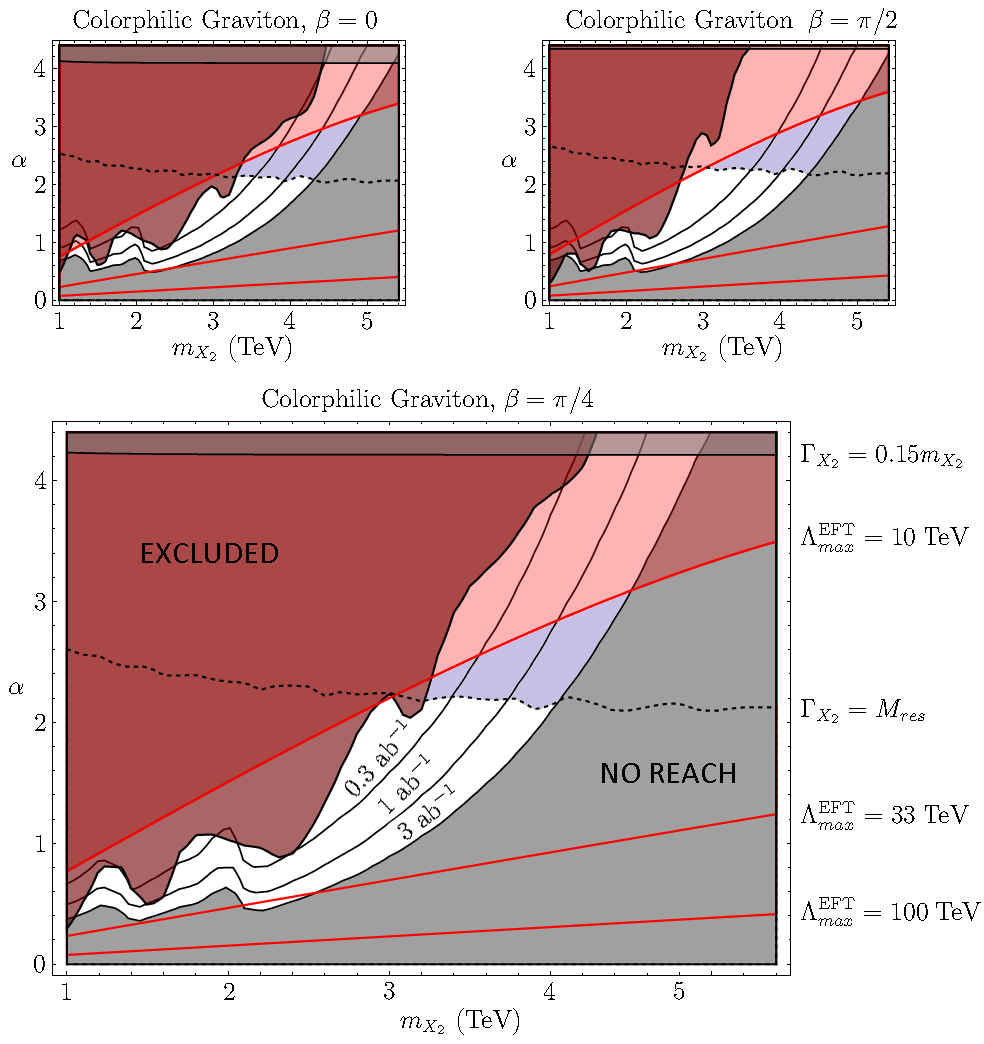}
\end{center}
\caption{Colorphilic graviton $(X_2)$ parameter space for various $\beta$ values. The translucent dark red region bounded by a thick black curve to the upper left is experimentally excluded at 95\% CL. The black curves denote points accessible with integrated luminosities of $\mathcal{L}_{int}=0.3,$ $1,$ and $3$ ab$^{-1}$ respectively; the gray dark region to the bottom-right is inaccessible even with $3$ ab$^{-1}$ of LHC-14 data. Points below the rising red diagonal lines respect tree-level unitarity up to $\Lambda_{max}^{\text{EFT}}=10$ TeV, $33$ TeV, and $100$ TeV from steeper to shallower lines respectively. The translucent red region violates tree-level unitarity below 10 TeV and is excluded from our analysis. The grayed region above the horizontal black line along the upper edge violates $\Gamma_{X_2}/m_{X_2}>0.15$. Points below the approximately-horizontal dashed $\Gamma_{X_2}=M_{res}$ curve have irresolvable widths and are excluded from $D_{col,X_2}$ analysis. For each $\beta$-value, this leaves the approximately triangular purple regions near $(m_{X_2},\alpha) = (3.8\text{ TeV},2.5)$ for $D_{col,X_2}$ analysis.}
\label{ParameterSpace}
\end{figure}

The mass resolution of the LHC is an approximately linear function of resonance mass $m_{R}$. We obtain the LHC dijet mass resolution function by interpolating resonance mass bin widths \cite{CMS2017}, and subsequently demand $\Gamma_{X_2} \geq M_{res}$. The approximately-horizontal dashed curve passing through the middle of each plot of Fig. \ref{ParameterSpace} denotes points for which $\Gamma_{X_2} = M_{res}$. Above this curve, colorphilic gravitons satisfy $\Gamma_{X_2} > M_{res}$ and possess resolvable widths at the LHC; below this curve, colorphilic gravitons satisfy $\Gamma_{X_2} < M_{res}$ and have irresolvable widths at the LHC.

\subsection{Summary of Available Parameter Space}
The regions of parameter space in Fig. \ref{ParameterSpace} are, in summary,
\begin{itemize}
\item {\bf Experimentally Excluded:} points within the translucent dark red region bounded by a thick black curve and located in the upper left of each plot are experimentally excluded at 95\% CL.
\item {\bf Experimentally Inaccessible:} points within the gray region to the bottom-right of each plot are deemed inaccessible at LHC-14, because they require more than $3$ ab$^{-1}$ worth of LHC-14 dijet channel data for discovery in the dijet channel.
\item {\bf Unitarity:} points within the translucent pale red region bounded by a red curve are excluded from our analysis because they violate unitarity below $\Lambda^{\text{EFT}}_{max}=10$ TeV.
\item {\bf Narrow Width Approximation:} points within the grayed region above the horizontal black line at each plot's upper edge satisfy $\Gamma_{X_2}/ m_{X_2} > 0.15$ and are excluded because the CMS and ATLAS narrow dijet resonance searches have poor sensitivity in this region \cite{NWA}.
\item {\bf Mass Resolution:} points below the approximately-horizontal dashed curve passing through the middle of each plot are excluded from the $D_{col,X_2}$ analysis because they possess unresolvable widths: $\Gamma_{X_2} \leq M_{res}$. Colorphilic gravitons with properties corresponding to points below this curve are still potentially discoverable at the LHC, but would have to be identified by means other than $D_{col,X_2}$.
\item {\bf  Color Discriminant Variable:} points in the approximately triangular purple regions of each plot are relevant to a color discriminant variable analysis.
\end{itemize}

The black curves correspond to the experimental reach for discovery of a colorphilic graviton at LHC-14 with luminosities $\mathcal{L}_{int}=0.3$, $1$, and $3$ ab$^{-1}$ from left-to-right, with parameter space below a given curve being inaccessible with only $\mathcal{L}_{int}$ worth of LHC-14 dijet channel data. The remaining white and purple regions in each Fig. \ref{ParameterSpace} subplot denote areas of parameter space accessible with $\mathcal{L}_{int}$ for the indicated $\beta$ value. As illustrated, LHC-14 has discovery reach for colorphilic gravitons with any $\beta$ value, even after taking unitarity constraints into account. Therefore, the colorphilic graviton $X_2$ is a new physics object that is generally relevant to searches in the LHC-14 dijet channel.

\vspace{1cm}
Several conclusions can be reached by comparing the various pieces of information overlapping within the panes of Fig. \ref{ParameterSpace}:

There is a reasonably large range of masses where a graviton can be discovered but a color discriminant variable analysis is inapplicable. This is because the discoverable parameter space corresponds to gravitons with relatively weak couplings and thus relatively small decay widths. There is a much smaller window of masses where the theory is consistent up to $\Lambda_{max}^{\text{EFT}}=$ 10 TeV yet the coupling is also large enough for the color discriminant analysis to work. Note additionally that this region quickly runs towards larger coupling parameter $\alpha$ with increasing graviton mass and consequently runs into conflict with the unitarity constraints. Among the different values of $\beta$, the color discriminant analysis applies for a slightly smaller graviton mass in the $\beta=\tfrac{\pi}{4}$ case than the other cases.

Given a fixed value of $\beta$, each value of $\Lambda^{\text{EFT}}_{max}$ generates a contour through parameter space. The $\Lambda^{\text{EFT}}_{max}$ contours corresponding to 10, 33, and 100 TeV are illustrated as red curves in Fig. \ref{ParameterSpace}. Note additionally the white and purple regions above the $\mathcal{L}_{int}=3$ ab$^{-1}$ curve: this is the region of parameter space where colorphilic gravitons are discoverable with $3$ ab$^{-1}$ of LHC-14 dijet data according to our criteria. As $\Lambda^{\text{EFT}}_{max}$ is continuously increased from $10$ TeV, the corresponding contour moves continuously downward, so that less of this discoverable region lies below the curve. There exists a value of $\Lambda^{\text{EFT}}_{max}$ for which no parameter space points are simultaneously discoverable with $\mathcal{L}_{int}=3$ ab$^{-1}$ LHC-14 data and consistent with our unitarity bounds up to $\Lambda^{\text{EFT}}_{max}$. This value of $\Lambda_{max}^{\text{EFT}}$ is $34$, $38$, and $36$ TeV for $\beta=0$, $\tfrac{\pi}{4}$, and $\tfrac{\pi}{2}$ respectively. Any colorphilic graviton arising from a theory consistent with larger values of $\Lambda_{max}^{\text{EFT}}$ would not be discoverable at LHC-14. Therefore, if a colorphilic graviton is discovered at the LHC, it necessarily implies additional new physics below about $30-40$ TeV.

The LHC-14 discoverable region relevant to a color discriminant variable analysis (each purple region) also corresponds to an upper limit on possible $\Lambda_{max}^{\text{EFT}}$ values. When $\mathcal{L}_{int}=3$ ab$^{-1}$, this uppermost value of $\Lambda_{max}^{\text{EFT}}$ is $15$, $14$, and $13$ TeV for $\beta=0$, $\tfrac{\pi}{4}$, and $\tfrac{\pi}{2}$ respectively. While theories of colorphilic gravitons with larger values of $\Lambda_{max}^{\text{EFT}}$ can be constructed, a color discriminant variable analysis would not be valid for such objects, because their widths would not be resolvable by the LHC detectors.

For ease of discussion during the color discriminant variable analysis in the next section, we denote by $\mathcal{P}_{\beta}[\mathcal{L}_{int}]$ the region in the $(m_{X_2},\alpha)$ plane of $X_2$ parameter space that is allowed by the above constraints ($\Lambda_{max}^{\text{EFT}}$ fixed at $10$ TeV) for a given luminosity $\mathcal{L}_{int}$ and $\beta$ value, and also generates dijet resonances with resolvable widths. These are exactly the purple regions of Figure 1. $\mathcal{P}_{\beta}[\mathcal{L}_{int}]$ is therefore a proper subset of the parameter space region where a colorphilic graviton is LHC-14 discoverable.

\section{Distinguishing $X_2$ from other Dijet Resonances}

Suppose a dijet resonance is observed at a mass $m_R$ and possesses a measurable color discriminant variable $D_{col,R}$. After accounting for experimental uncertainties, that measurement may or may not be consistent with the value of $D_{col}$ predicted by a model. In this way, experiments can immediately eliminate any class of models inconsistent with the measured $D_{col,R}$ as an explanation of that observed dijet resonance. The utility of the color discriminant variable analysis hinges on this capability.

The measurable $D_{col,R}$ values at a fixed $m_R=m_{X_2}$ that are consistent with the colorphilic graviton model depend on the available parameter space (described in Section 3) as well as the uncertainties of the measurements involved. In what follows, we review the statistical and systematic uncertainties relevant to the color discriminant variable; plot the color discriminant variable (with uncertainties) for the colorphilic graviton; and show in detail how the color discriminant variable can be used to distinguish colorphilic gravitons from other models.

Error propagation of the color discriminant variable is detailed in Ref \cite{Atre:2013mja} and summarized here. All uncertainties are modeled as Gaussian. The uncertainty of $\log_{10}D_{col,R}$ is related to the relative uncertainties of the dijet cross-section, mass, and decay width according to,
\begin{align}
\dfrac{\Delta \left[\log_{10}D_{col,R}\right]}{0.434}=\dfrac{\Delta D_{col,R}}{D_{col,R}}=\left(\dfrac{\Delta\sigma_{Rjj}}{\sigma_{Rjj}}\right)\oplus\left(3\dfrac{\Delta m_R}{m_R}\right)\oplus\left(\dfrac{\Delta \Gamma_R}{\Gamma_R}\right)
\end{align}
The symbol $\oplus$ denotes addition in quadrature. The relative error of the dijet cross-section is,
\begin{align}
\dfrac{\Delta \sigma_{Rjj}}{\sigma_{Rjj}}=\dfrac{1}{\sqrt{N_{5\sigma}}}\oplus \epsilon_{\sigma\text{ sys}}
\end{align}
where $N_{5\sigma}$ denotes the number of events necessary to make a $5\sigma$ discovery and $\epsilon_{\sigma SYS}$ is the dijet cross-section's systematic uncertainty. The relative uncertainty of the dijet mass is,
\begin{align}
\dfrac{\Delta m_R}{m_R}=\left[\dfrac{1}{\sqrt{N_{5\sigma}}}\left(\dfrac{\sigma_\Gamma}{m_R}\oplus\dfrac{M_{res}}{m_R}\right)\right]\oplus \left(\dfrac{\Delta M}{M}\right)_{\text{JES}}
\end{align}
where $\sigma_\Gamma\simeq \Gamma_R/2.35$ is the (Gaussian) standard deviation of the resonance's intrinsic width. $M_{res}$ denotes the previously-mentioned dijet mass resolution of the experiment and $(\Delta M_{JES}/M)$ is mass measurement uncertainty due to jet energy scale uncertainties. The relative uncertainty of the dijet resonance total decay width is,
\begin{align}
\dfrac{\Delta \Gamma_R}{\Gamma_R} =\sqrt{\dfrac{1}{2(N_{5\sigma}-1)}\left[1+\left(\dfrac{M_{res}}{\sigma_\Gamma}\right)^2\right]^2+\left(\dfrac{M_{res}}{\sigma_\Gamma}\right)^4\left(\dfrac{\Delta M_{res}}{M_{res}}\right)^2}
\end{align}
where $\Delta M_{res}/M_{res}$ is the relative uncertainty of the dijet mass resolution. The various uncertainties entering this calculation have been estimated using experimental data, and are summarized as follows:
\begin{itemize}
\item $\epsilon_{\sigma\text{ sys}}$ is an approximately linear function of $m_{X_2}$, extracted from \cite{Gumus:2006mxa}.
 $$\epsilon_{\sigma\text{ sys}}(1\text{ TeV})=0.24\hspace{35 pt}\epsilon_{\sigma\text{ sys}}(6\text{ TeV})=0.42$$
\item $M_{res}$ is an approximately linear function of $m_{X_2}$ obtained as described in Section 3.5, by interpolating bin widths \cite{CMS2017}.
$$M_{res}(1\text{ TeV})=0.057\text{ TeV}\hspace{35 pt}M_{res}(6\text{ TeV})=0.23\text{ TeV}$$
\item $(\Delta M/M)_{JES} = 0.013$ \cite{Chatrchyan:2013qha}.
\item $\Delta M_{res}/M_{res} = 0.1$ \cite{Chatrchyan:2013qha}.
\end{itemize}

\begin{figure}[p]
\begin{center}
\includegraphics[scale=0.65]{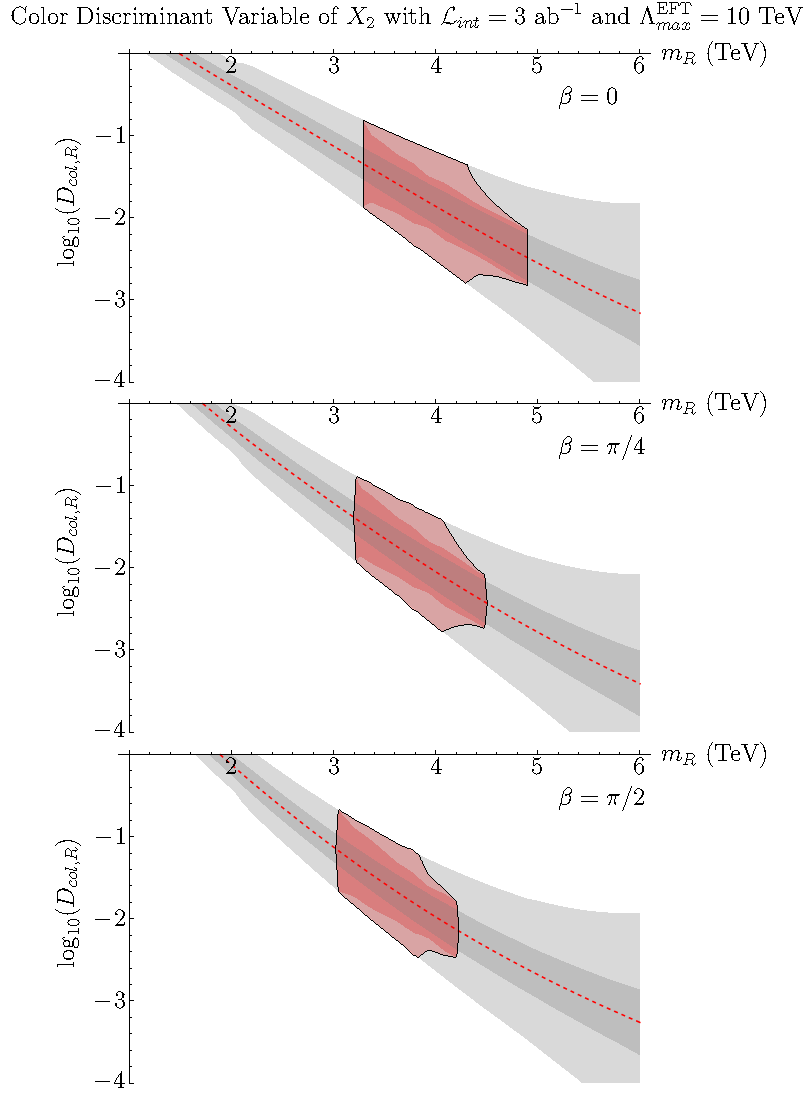}
\end{center}
\caption{The color discriminant variable and its uncertainties for the colorphilic graviton $X_2$. $D_{col,X_2}$ is plotted as a dashed red curve for each value of $\beta$ considered.  In the long shaded regions: dark gray denotes $D_{col,R}$ values within $1\sigma$ of $D_{col,X_2}$ when uncertainties are calculated with $\Gamma_{X_2}/M_{X_2}=0.15$. Light gray denotes $D_{col,R}$ values within $1\sigma$ of $D_{col,X_2}$ when uncertainties are calculated with fixed $\Gamma_{X_2}=M_{res}$, the mass resolution of the LHC. In the truncated shaded regions near 3.8 TeV: solid (faded) red denotes $D_{col,R}$ values within $1\sigma$ of $D_{col,X_2}$ when uncertainties are calculated with the minimum (maximum) uncertainty available in the purple $\mathcal{L}_{int}=3$ ab$^{-1}$ parameter space from Fig. \ref{ParameterSpace}. A black border outlines the faded red region.} 
\label{CDVLGr}
\end{figure}

We analyze the $\mathcal{L}_{int}=3$ ab$^{-1}$ parameter spaces $\mathcal{P}_\beta[3\text{ ab}^{-1}]$ The results of the color discriminant variable analysis are summarized in Fig. \ref{CDVLGr}; the plots within the figure correspond to $\beta=0$, $\tfrac{\pi}{4}$, and $\tfrac{\pi}{2}$ from top to bottom respectively.

Each plot within Fig. \ref{CDVLGr} contains a dashed red curve that traces out the theoretical color discriminant variable $D_{col,X_2}$ as a function of $m_{X_2}$ with $s=(14\text{ TeV})^2$. By construction, $D_{col,X_2}$ only depends on $m_{X_2}$; at tree level, any direct $\alpha$ dependence cancels out. However, secondary $\alpha$ dependence lingers in the experimental uncertainty of a $D_{col,R}$ measurement. The various bands we have plotted in $(m_{R},D_{col,R})$ space correspond to values for which a measurement of a dijet resonance with mass $m_{R}$ and color discriminant variable $D_{col,R}$ would be consistent with the theoretical $D_{col,X_2}$, with different colors signifying different conditions:
\begin{itemize}
\item {\bf Darker Gray:} Values of $(m_{R},D_{col,R})$ within $1\sigma$ of the theoretical $D_{col,X_2}$, where the uncertainty has been obtained by ignoring all constraints and fixing $\Gamma_{X_2}/m_{X_2} = 0.15$, the boundary of the narrow width approximation.
\item {\bf Lighter Gray:} Values of $(m_{R},D_{col,R})$ within $1\sigma$ of the theoretical $D_{col,X_2}$, where the uncertainty has been obtained by ignoring all constraints and fixing $\Gamma_{X_2} = M_{res}$, the boundary determined by the mass resolution of the LHC.
\item {\bf Solid (Faded) Red:}  Values of $(m_{R},D_{col,R})$ within $1\sigma$ of the theoretical $D_{col,X_2}$, where the uncertainty is set to the minimum (maximum) uncertainty available in $\mathcal{P}_\beta[3\text{ ab}^{-1}]$, as plotted in the purple region of Fig. \ref{ParameterSpace}. At any given $m_{X_2}$, this corresponds to the largest (smallest) width available for that given mass in $\mathcal{P}_\beta[3\text{ ab}^{-1}]$. A black border outlines the faded red region.
\end{itemize}

The usefulness of the color discriminant variable emerges from comparing the different regions of $(m_R,D_{col,R})$ space that various dijet resonances $R$ occupy. For example, consider the leptophobic $Z^\prime$ particle from Ref \cite{Atre:2013mja}. Like $X_2$, the leptophobic $Z^\prime$ possesses a parameter space restricted by the NWA, mass resolution of the detector, exclusion limits, and $\mathcal{L}_{int}$ discovery prospects, yielding an available parameter space $\mathcal{P}_{Z^\prime}[\mathcal{L}_{int}]$. An updated color discriminant variable analysis on the leptophobic $Z^\prime$ is summarized in Appendix A. Previous works demonstrated that this $Z^\prime$ is well-separated from colorons, excited quarks, and diquarks in $(m_R,D_{col,R})$ space \cite{Atre:2013mja},\cite{Simmons:2015gna}-\cite{Chivukula:2015zma}. As a result, a measured $D_{col,R}$ consistent with a leptophobic $Z^\prime$ is necessarily inconsistent with coloron, excited quark, and diquark models.

\begin{figure}[p]
\begin{center}
\includegraphics[scale=0.65]{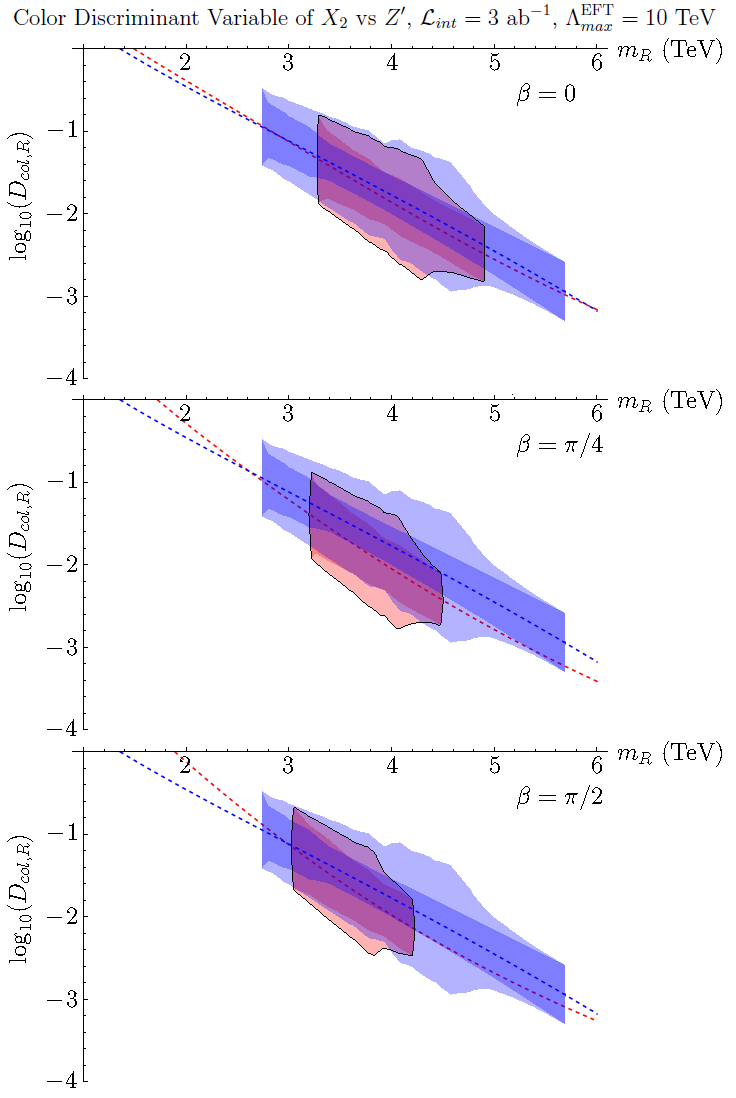}
\end{center}
\caption{Comparison of the color discriminant variable and its uncertainties for the colorphilic graviton $X_2$ versus the leptophobic $Z^\prime$. $D_{col,X_2}$ is plotted as a dashed red curve for each value of $\beta$ considered. Solid (faded) red denotes $D_{col,R}$ values within $1\sigma$ of $D_{col,X_2}$ when uncertainties are calculated with the minimum (maximum) uncertainty available in the $\mathcal{L}_{int}=3$ ab$^{-1}$ parameter space from Fig. 1; this reproduces the red regions of Figure 2. A black border outlines the faded red region. The color discriminant variable $D_{col,Z^\prime}$ of the leptophobic $Z^\prime$ is plotted as a dashed blue curve. Solid (faded) blue denotes $D_{col,R}$ values within $1\sigma$ of $D_{col,Z^\prime}$ when uncertainties are calculated with the minimum (maximum) uncertainty available in the $\mathcal{L}_{int}=3$ ab$^{-1}$ parameter space for $Z^\prime$.}
\label{MultiCDV}
\end{figure}

Fig.~\ref{MultiCDV} shows how the colorphilic graviton $X_2$ compares to the leptophobic $Z^\prime$ in $(m_R,D_{col,R})$ space for $\beta=0$, $\tfrac{\pi}{2}$, and $\tfrac{\pi}{4}$ from top to bottom respectively. As in Fig. \ref{CDVLGr}, $\mathcal{L}_{int}$ is fixed at $3$ ab$^{-1}$ and the colorphilic graviton is illustrated in red. The leptophobic $Z^\prime$ is illustrated in blue. The dashed blue line plots $D_{col,Z^\prime}$ and the regions are colored according to
\begin{itemize}
\item {\bf Solid (Faded) Blue:}  Values of $(m_{R},D_{col,R})$ within $1\sigma$ of the theoretical $D_{col,R}$, where the uncertainty is set to the minimum (maximum) uncertainty available in $\mathcal{P}_{Z^\prime}(3\text{ ab}^{-1})$. At any given $m_{R}$, this corresponds to the largest (smallest) width available for that given mass in $\mathcal{P}_{Z^\prime}(3\text{ ab}^{-1})$.
\end{itemize}
We use the same procedures and sources for the $Z^\prime$ data as the $X_2$ data, and utilize procedures identical to those described in Section 3.

We immediately see from Fig. \ref{MultiCDV} that most regions of $(m_{R},D_{col,R})$ space occupied by the colorphilic graviton are already occupied by the leptophobic $Z^\prime$. The $\beta=\tfrac{\pi}{4}$ models are potentially distinguishable from $Z^\prime$ models at resonant masses $m_{R}\sim 3.5\text{ TeV}$, although even these points remain within $2\sigma$ of the $Z^\prime$ models. Meanwhile, effectively all resolvable dijet resonances consistent with $\beta=0$ colorphilic graviton models are also consistent with leptophobic $Z^\prime$ models.

There is, however, a significant amount of $(m_R,D_{col,R})$ space where a resolvable dijet resonance would be consistent with a leptophobic $Z^\prime$ model but is inconsistent with $X_2$ models. This is because $X_2$ is more restricted in its range of available masses by unitarity constraints. For instance, for all values of $\beta$, colorphilic gravitons with masses below $m_{X_2}\lesssim 3\text{ TeV}$ are either excluded, violate unitarity below $10$ TeV, or possess irresolvable widths. As such, a resolvable dijet resonance with $m_{R}\lesssim 3\text{ TeV}$ is unlikely to be explained by $X_2$, but may possibly be described by a leptophobic $Z^\prime$. Similarly, a resolvable dijet resonance with $m_{R}\gtrsim 5\text{ TeV}$ is unlikely to be described by a colorphilic graviton.

Although the significant overlap between leptophobic $Z^\prime$ and $X_2$ models means many resolvable dijet resonances could be equally well described by either model, it also means colorphilic gravitons can be distinguished from coloron, excited quark, and diquark models just like the leptophobic $Z^\prime$. If a $D_{col,R}$ measurement falls in the regime where the colorphilic graviton and $Z^\prime$ overlap, then a detailed angular distribution analysis of the decay products will be required to determine the spin of the resonance and thereby select the appropriate model.

\section{Summary and Conclusions}
The LHC dijet channel is one of the most powerful tools of modern physics, capable of exploring previously-unseen physical regimes. If a dijet resonance $R$ is discovered, determining its origin will be of intense interest. Our work here has demonstrated that the LHC might yet discover a dijet resonance consistent with a ``colorphilic graviton" $X_2$, a massive spin-2 particle that couples to the quark and gluon stress-energy tensors.

Analyzing the phenomenology of the $X_2$ state requires careful consideration of unitarity constraints: because $X_2$ couples to $q\overline{q}$ and $gg$ states via dimension-5 operators, the colorphilic graviton generically violates tree-level unitarity. Colorphilic graviton models are parameterized by the particle's mass $m_{X_2}$, the overall coupling strength $\alpha$, a measure of the relative coupling strength of $X_2$ to the quark vs gluon stress-energy tensors $\beta$, and the scale $\Lambda_{max}^{\text{EFT}}$ up to which the $X_2$ model respects tree-level unitarity.

The parameter space of $X_2$ relevant to collider searches is constrained by unitarity considerations, application of the narrow width approximation, experimental exclusions, and ability to be discovered with integrated luminosity $\mathcal{L}_{int}$. The region of $\alpha$ vs $m_{X_2}$ parameter space that survives these constraints is illustrated in white and purple in Fig. \ref{ParameterSpace} for $\mathcal{L}_{int}=0.3$, $1$, $3\text{ ab}^{-1}$, and $\beta=0$, $\tfrac{\pi}{4}$, $\tfrac{\pi}{2}$. For every value of $\beta$, we found that there does exist a region of parameter space in which the LHC could discover a $X_2$-originating dijet resonance with $3$ ab$^{-1}$ of data. Moreover, across the entire range of $\beta$ values, we see that there is an upper limit of about $5\text{ TeV}$ on the mass of a discoverable $X_2$ resonance. For self-consistency, the value of $\Lambda^{\text{EFT}}_{max}$ must be at least this large. Accordingly, we set $\Lambda_{max}^{\text{EFT}}=10\text{ TeV}$ for this analysis. Due to the combined effect of discovery prospects and unitarity constraints, we also found that the discovery of a colorphilic graviton would necessarily imply the presence of additional new physics below about 30-40 TeV. 

While a dijet resonance $R$ consistent with a colorphilic graviton could be discovered at the LHC, such a resonance might also initially be consistent with other models. The color discriminant variable $D_{col,R}$ provides a quick means of eliminating potential dijet resonance models \cite{Atre:2013mja}. $D_{col,R}$ is constructed from quantities immediately measurable after the discovery of a resolvable resonance: the mass of the resonance $m_R$, the total dijet cross section at the peak of the resonance $\sigma_{Rjj}$, and the width of the resonance $\Gamma_R$. The color discriminant variable has proven useful when applied to leptophobic $Z^\prime$, coloron, excited quarks, and diquark models \cite{Atre:2013mja},\cite{Simmons:2015gna}-\cite{Chivukula:2015zma}. 

In applying the color discriminant variable analysis to $X_2$, we fixed $\mathcal{L}_{int}=3\text{ ab}^{-1}$ to get an idea of which resolvable colorphilic gravitons might be observed in the LHC dijet channel. Our subsequent analysis revealed a limited range of color discriminant variable $D_{col,R}$ values consistent with a colorphilic graviton $X_2$ to within $1\sigma$ of experimental uncertainties, as summarized in Fig. \ref{CDVLGr}. Furthermore, the region of $(m_R,D_{col,R})$ space consistent with $X_2$ is largely shared with the leptophobic $Z^\prime$, so that resolvable dijet resonances consistent with $X_2$ are also consistent with $Z^\prime$. This is demonstrated in Fig. \ref{MultiCDV}. Therefore, just as $D_{col,R}$ is able to discriminate $Z^\prime$ resonances from coloron, excited quark, and diquark resonances, it can likewise tell an $X_2$ apart from those other states. Should a $D_{col,R}$ measurement fall in the values applicable to both the colorphilic graviton and $Z^\prime$, additional analyses would be required to distinguish the models. For example, it might be possible to recognize the three- and four-body decay modes of the colorphilic graviton with sufficient luminosity. More generally, the spin of the new resonance could be determined via a detailed angular analysis of its decay products. 

We look forward to future opportunities to apply the results of our analysis to newly discovered dijet resonances.

\section*{Acknowledgments}

We thank Kirtimaan Mohan for useful conversations. The work of. R.S.C., D.F., and  E.H.S. was supported by the National Science Foundation under Grant PHY-1519045. 

\section*{A. Parameter Space of Leptophobic $Z^\prime$}
The (flavor-universal) leptophobic $Z^\prime$ is a massive spin-1 particle with the following interaction Lagrangian:
\begin{align}
\mathcal{L}_{Z^\prime,eff} = igZ_\mu^\prime\sum_{i}\overline{q}_i\gamma^\mu (g_L P_L + g_R P_R)q_i
\end{align}
where $i$ sums over the SM quarks, $g$ is the weak coupling, and $P_{L(R)}$ is the left (right) projection operator \cite{Atre:2013mja}. We calculate quantities in the large $Z^\prime$ mass limit, so the couplings only occur in the $|g_L|^2+|g_R|^2$ combination.

\begin{figure}[p]
\begin{center}
\includegraphics[scale=0.55]{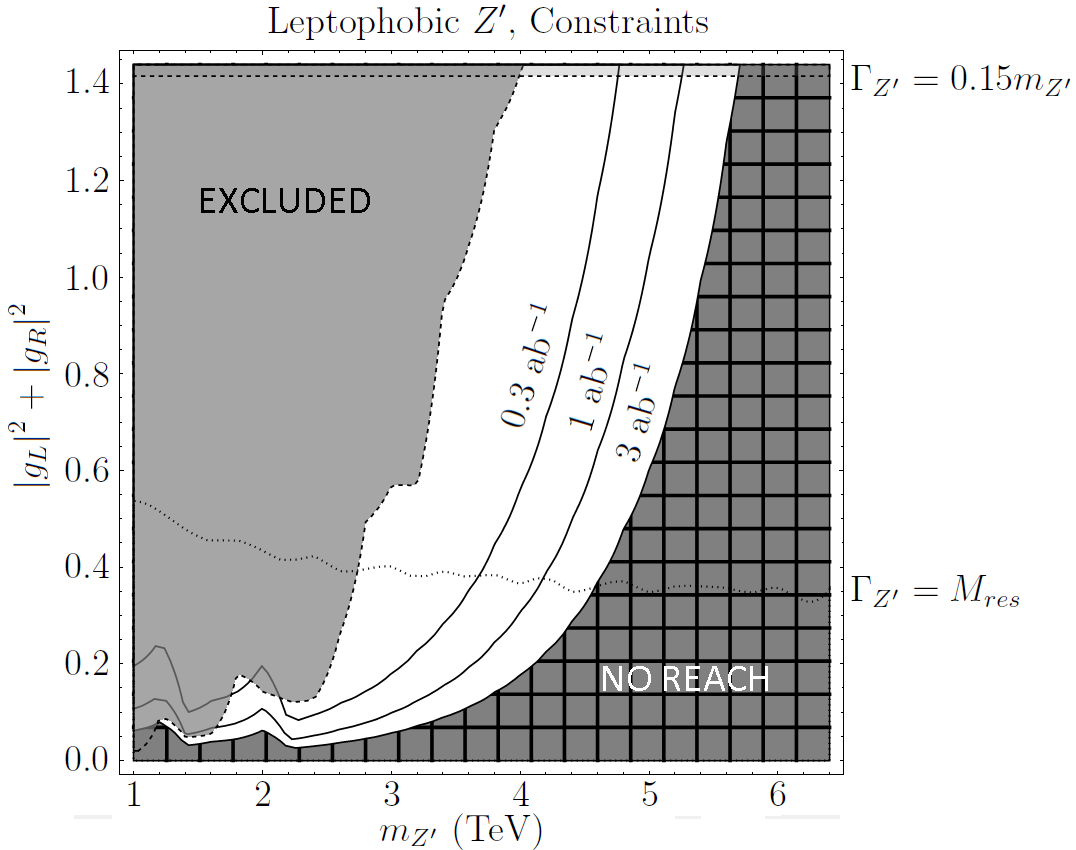}
\end{center}
\caption{Leptophobic $Z^\prime$ parameter space. The lightly-grayed region above the horizontal dashed line along the upper edge violates $\Gamma_{Z^\prime}/m_{Z^\prime}>0.15$.   Points below the approximately-horizontal dotted $\Gamma_{Z^\prime}=M_{res}$ curve have irresolvable widths and are excluded from $D_{col,Z^\prime}$ analysis. The translucent dark gray region bounded by a dashed line to the upper left is experimentally excluded to 95\%. The black curves denote points accessible with integrated luminosities of $\mathcal{L}_{int}=0.3,$ $1,$ and $3$ ab$^{-1}$ respectively; the cross-hatched dark region to the bottom-right is inaccessible even with $3$ ab$^{-1}$ of LHC-14 data. This leaves the white regions above the $\Gamma_{Z^\prime}=M_{res}$ curve for $D_{col,Z^\prime}$ analysis.}
\label{ParameterSpaceZp}
\end{figure}

As with the colorphilic graviton, the parameter space of the leptophobic $Z^\prime$ is constrained by experimental exclusions, application of the narrow width approximation, discovery prospects, and the requirement of a resolvable decay width. We utilize the CMS $95\%$ exclusion data from \cite{CMS2017} and the projected $5\sigma$ discovery prospects from \cite{Gumus:2006mxa}, and apply them via the methods described in Section 3. Fig. \ref{ParameterSpaceZp} summarizes the parameter space that survives this procedure.

The color discriminant variable analysis proceeds identically to the analysis of $X_2$, utilizing the same experimental uncertainties and application methods. Fig. \ref{MultiCDV} illustrates the collection of $(m_R,\log_{10}D_{col,R})$ measurements consistent with the leptophobic $Z^\prime$ model for $\mathcal{L}_{int}=3\text{ ab}^{-1}$ worth of LHC- dijet channel data. Specifically, the solid (faded) blue denotes $D_{col,R}$ values within $1\sigma$ of the theoretical $D_{col,Z^\prime}$ when uncertainties are calculated with the minimum (maximum) uncertainty available in the $\mathcal{L}_{int}=3$ ab$^{-1}$ parameter space for the leptophobic $Z^\prime$.

\section*{B. Tree-Level Partial-Wave Amplitude Details}
We now outline the construction of the full scattering matrix in the limit $m_t\ll \sqrt{\hat{s}}$. For $s$-channel $2\rightarrow X_2\rightarrow 2$ scattering of massless fermions $f_1\overline{f}_1$ to massless fermions $f_2\overline{f}_2$,
\begin{align}
\mathcal{M}_{f_1\overline{f}_1\rightarrow f_2\overline{f}_2} = -\dfrac{c_\beta^2\alpha^2\hat{s}^2}{16m_{X_2}^2(\hat{s}-m_{X_2}^2)}\left(\hspace{-5 pt}\begin{tabular}{ c c }
$d^2_{+1,+1}$ & $d^2_{+1,-1}$ \\
$d^2_{-1,+1}$ & $d^2_{-1,-1}$
\end{tabular}\hspace{-5 pt}\right)
\end{align}
where $d^J_{m_1,m_2}(\cos\theta)$ denotes the Wigner $d$-functions \cite{PDG} (where we set $\phi=0$ without loss of generality). Similarly, for massless fermions to massless gauge bosons and vice-versa,
\begin{align}
\mathcal{M}_{f_i\overline{f}_i\rightarrow \gamma\overline{\gamma}}  &=-\dfrac{s_\beta c_\beta \alpha^2 \hat{s}^2}{8m_{X_2}^2(\hat{s}-m_{X_2}^2)}\left(\hspace{-5 pt}\begin{tabular}{ c c }
$d^2_{+1,+2}$ & $d^2_{+1,-2}$ \\
$d^2_{-1,+2}$ & $d^2_{-1,-2}$ 
\end{tabular}\hspace{-5 pt}\right)\\
\mathcal{M}_{\gamma\overline{\gamma}\rightarrow f_i\overline{f}_i}&=-\dfrac{s_\beta c_\beta \alpha^2 \hat{s}^2}{8m_{X_2}^2(\hat{s}-m_{X_2}^2)}\left(\hspace{-5 pt}\begin{tabular}{ c c }
$d^2_{+2,+1}$ & $d^2_{+2,-1}$ \\
$d^2_{-2,+1}$ & $d^2_{-2,-1}$
\end{tabular}\hspace{-5 pt}\right)
\end{align}
Finally, for massless gauge bosons to massless gauge bosons,
\begin{align}
\mathcal{M}_{\gamma\overline{\gamma}\rightarrow \gamma\overline{\gamma}} =-\dfrac{s_\beta^2 \alpha^2 \hat{s}^2}{4m_{X_2}^2(\hat{s}-m_{X_2}^2)}\left(\hspace{-5 pt}\begin{tabular}{ c c c c }
$d^2_{+2,+2}$ & $d^2_{+2,-2}$ \\
$d^2_{-2,+2}$ & $d^2_{-2,-2}$
\end{tabular}\hspace{-5 pt}\right)
\end{align}
The rows (columns) label initial (final) state helicity combinations and are organized from top-to-bottom (left-to-right) as follows: $(+h,-h)$, $(-h,+h)$, where $h=\tfrac{1}{2}$ for the fermion-antifermion states and $h=1$ for the massless vector boson states. The $(+h,+h)$ and $(-h,-h)$ combinations yield vanishing matrix elements in the massless limit. To obtain the full scattering matrix, we piece together matrices of these sort for every flavor and color of particles that couple to the colorphilic graviton. Columns with outgoing gluons should be divided by two to avoid double-counting identical particles. This final transition matrix possesses $(2\cdot 6\cdot 3 + 2\cdot 8)^2 = 52^2$ elements, each of which is converted to a partial-wave amplitude via the following formula:
\begin{align}
\mathcal{M}_{2_{(a,b)}\rightarrow X_2\rightarrow 2_{(c,d)}} = 80\pi \mathcal{A}_{2_{(a,b)}\rightarrow X_2 \rightarrow 2_{(c,d)}} d^2_{b-a,d-c} 
\end{align}
The partial-wave unitarity constraint is then that every eigenvalue $a_2$ of the partial wave amplitude matrix $\mathcal{A}$ satisfy $|\mathfrak{R}[a_2]|\leq \tfrac{1}{2}$. The resulting matrix $\mathcal{A}$ of partial-wave amplitudes is highly redundant, and possesses only one nonzero eigenvalue $a_2$, such that,
\begin{align}
a_2 &= -\dfrac{\alpha^2 \hat{s}^2}{80\pi m_{X_2}^2(\hat{s}-m_{X_2}^2)}\left[\dfrac{c_\beta^2}{16}\cdot 2\cdot 6\cdot 3 +\dfrac{1}{2}\cdot  \dfrac{s_\beta^2}{4} \cdot 2\cdot 8 \right] = -\dfrac{(8+c_\beta^2)\alpha^2 \hat{s}^2}{320\pi m_{X_2}^2(\hat{s}-m_{X_2}^2)}
\end{align}
where $s_{\beta}^2$ has been eliminated via the fundamental trigonometric identity. Note the pole at $\hat{s}=m_{X_2}^2$, which we generally expect. For convenience, we define the dimensionless combinations,
\begin{align}\label{eq:simpcombo}
\hat{\mathfrak{s}} \equiv \dfrac{\hat{s}}{m_{X_2}^2} \hspace{75 pt} \mathfrak{g} \equiv \dfrac{(8+c_\beta^2)\alpha^2}{320\pi}
\end{align}
We have defined $\mathfrak{g}$ intentionally such that $\mathfrak{g}= \Gamma_{X_2}/m_{X_2}$. Partial-wave unitarity subsequently demands,
\begin{align}
\tfrac{1}{2} \geq |\mathfrak{R}[a_2]| \hspace{20 pt}\implies\hspace{20 pt} |\hat{\mathfrak{s}}-1| \geq 2\hspace{2 pt}\mathfrak{g}\hspace{2 pt}\hat{\mathfrak{s}}^2
\end{align}
There is a maximum value of $\hat{\mathfrak{s}}$ that respects this constraint given a specific colorphilic graviton model. Specifically, for a given $\mathfrak{g}$, unitarity requires,
\begin{align}\label{eq:smax}
\hat{\mathfrak{s}} < \hat{\mathfrak{s}}_{max} \equiv \dfrac{1}{4\mathfrak{g}}\left[1+\sqrt{1-8\mathfrak{g}}\right]
\end{align}
For $\mathfrak{g} > \tfrac{1}{8}$, no value of $\hat{\mathfrak{s}}$ solves Eq. (\ref{eq:smax}), putting an upper limit on width of the colorphilic graviton that can be consistent with unitarity. Section 3.4 will show this constraint is even stronger than the constraint due to the narrow width approximation.

We define $\Lambda^{\text{EFT}}_{max}$ such that $\hat{\mathfrak{s}}_{max}\equiv (\Lambda^{\text{EFT}}_{max}/m_{X_2})^2$. Given a specific instance of a colorphilic graviton model with parameters $m_{X_2*}$, $\alpha_*$, and $\beta_*$, tree-level unitarity is respected only for partonic center-of-momentum energies $\sqrt{\hat{s}}$ that satisfy,
\begin{align}\label{eq:unicon}
\sqrt{\hat{s}} < \Lambda_{max}^{\text{EFT}}\hspace{15 pt}\text{ where }\hspace{15 pt}\Lambda^{\text{EFT}}_{max} = m_{X_2*}\left[\dfrac{80\pi}{(8+c_{\beta_*}^2)\alpha_*^2}\left(1+\sqrt{1-(8+c_{\beta_*}^2)\dfrac{\alpha_*^2}{40\pi}}\right)\right]^{1/2}
\end{align}
NLO corrections can play a large role in certain spin-2 production channels \cite{Artoisenet:2013puc}. These corrections become relevant when the $p_T$ of the produced spin-2 object is on the order of (or larger than) the graviton mass $m_{X_2}$. The present article considers resonant production of a graviton with a several TeV mass near threshold, and at low $p_T$. In this regime, those NLO corrections do not dominate \cite{Das:2016pbk}.

Additional NLO corrections may originate from the use of non-universal couplings. When $\kappa_q\neq \kappa_g$ and $\hat{s}\gg m_{X_2}$, the production cross-section at NLO contains a term proportional to $\hat{s}^3(\kappa_q-\kappa_g)^2/m_{X_2}^6$. However, the present article is restricted to models when the partonic quantity $\sqrt{\hat{s}}$ is never significantly larger than the graviton mass $m_{X_2}$. Therefore, the tree-level analysis is adequate for establishing a region of $X_2$ parameter space where the color discriminant variable may be applicable.


\begin{thebibliography}{99}
%\cite{CMS2017}
\bibitem{CMS2017} 
  CMS Collaboration [CMS Collaboration],
  %``Searches for dijet resonances in pp collisions at $\sqrt{s}=13~\mathrm{TeV}$ using data collected in 2016.,''
  CMS-PAS-EXO-16-056.
  %%CITATION = CMS-PAS-EXO-16-056;%%
  %11 citations counted in INSPIRE as of 31 Jul 2017
  
%\cite{ATLAS2017}
\bibitem{ATLAS2017} 
  M.~Aaboud {\it et al.} [ATLAS Collaboration],
  %``Search for new phenomena in dijet events using 37 fb$^{-1}$ of $pp$ collision data collected at $\sqrt{s}=$13 TeV with the ATLAS detector,''
  arXiv:1703.09127 [hep-ex].
  %%CITATION = ARXIV:1703.09127;%%
  %21 citations counted in INSPIRE as of 31 Jul 2017
  
  %\cite{Randall:1999ee}
\bibitem{Randall:1999ee} 
  L.~Randall and R.~Sundrum,
  %``A Large mass hierarchy from a small extra dimension,''
  Phys.\ Rev.\ Lett.\  {\bf 83}, 3370 (1999)
  doi:10.1103/PhysRevLett.83.3370
  [hep-ph/9905221].
  %%CITATION = doi:10.1103/PhysRevLett.83.3370;%%
  %7595 citations counted in INSPIRE as of 31 Jul 2017
  
%\cite{Artoisenet:2013puc}
\bibitem{Artoisenet:2013puc} 
  P.~Artoisenet {\it et al.},
  %``A framework for Higgs characterisation,''
  JHEP {\bf 1311}, 043 (2013)
  doi:10.1007/JHEP11(2013)043
  [arXiv:1306.6464 [hep-ph]].
  %%CITATION = doi:10.1007/JHEP11(2013)043;%%
  %109 citations counted in INSPIRE as of 18 Apr 2017


%\cite{Martini:2016ahj}
\bibitem{Martini:2016ahj} 
  A.~Martini, K.~Mawatari and D.~Sengupta,
  %``Diphoton excess in phenomenological spin-2 resonance scenarios,''
  Phys.\ Rev.\ D {\bf 93}, no. 7, 075011 (2016)
  doi:10.1103/PhysRevD.93.075011
  [arXiv:1601.05729 [hep-ph]].
  %%CITATION = doi:10.1103/PhysRevD.93.075011;%%
  %33 citations counted in INSPIRE as of 21 Feb 2017
  
  %\cite{Han:2015cty}
\bibitem{Han:2015cty} 
  C.~Han, H.~M.~Lee, M.~Park and V.~Sanz,
  %``The diphoton resonance as a gravity mediator of dark matter,''
  Phys.\ Lett.\ B {\bf 755}, 371 (2016)
  doi:10.1016/j.physletb.2016.02.040
  [arXiv:1512.06376 [hep-ph]].
  %%CITATION = doi:10.1016/j.physletb.2016.02.040;%%
  %145 citations counted in INSPIRE as of 31 Jul 2017
  
%\cite{Kim:2001rc}
\bibitem{Kim:2001rc} 
  C.~S.~Kim, J.~D.~Kim and J.~H.~Song,
  %``Muon anomalous magnetic moment (g-2)(muon) and the Randall-Sundrum model,''
  Phys.\ Lett.\ B {\bf 511}, 251 (2001)
  doi:10.1016/S0370-2693(01)00635-9
  [hep-ph/0103127].
  %%CITATION = doi:10.1016/S0370-2693(01)00635-9;%%
  %49 citations counted in INSPIRE as of 31 Jul 2017
  
%\cite{Gao:2009eq}
\bibitem{Gao:2009eq} 
  J.~Gao, C.~S.~Li, X.~Gao and J.~J.~Zhang,
  %``Signature of Large Extra Dimensions from Z boson pair production at the CERN Large Hadron Collider,''
  Phys.\ Rev.\ D {\bf 80}, 016008 (2009)
  doi:10.1103/PhysRevD.80.016008
  [arXiv:0903.2551 [hep-ph]].
  %%CITATION = doi:10.1103/PhysRevD.80.016008;%%
  %9 citations counted in INSPIRE as of 31 Jul 2017

%\cite{Han:2004wt}
\bibitem{Han:2004wt} 
  T.~Han and S.~Willenbrock,
  %``Scale of quantum gravity,''
  Phys.\ Lett.\ B {\bf 616}, 215 (2005)
  doi:10.1016/j.physletb.2005.04.040
  [hep-ph/0404182].
  %%CITATION = doi:10.1016/j.physletb.2005.04.040;%%
  %41 citations counted in INSPIRE as of 22 Feb 2017

  %\cite{Atre:2013mja}
\bibitem{Atre:2013mja} 
  A.~Atre, R.~S.~Chivukula, P.~Ittisamai and E.~H.~Simmons,
  %``Distinguishing Color-Octet and Color-Singlet Resonances at the Large Hadron Collider,''
  Phys.\ Rev.\ D {\bf 88}, 055021 (2013)
  doi:10.1103/PhysRevD.88.055021
  [arXiv:1306.4715 [hep-ph]].
  %%CITATION = doi:10.1103/PhysRevD.88.055021;%%
  %11 citations counted in INSPIRE as of 22 Feb 2017
  
  %\cite{Chivukula:2014pma}
\bibitem{Chivukula:2014pma} 
  R.~Sekhar Chivukula, E.~H.~Simmons and N.~Vignaroli,
  %``Distinguishing dijet resonances at the LHC,''
  Phys.\ Rev.\ D {\bf 91}, no. 5, 055019 (2015)
  doi:10.1103/PhysRevD.91.055019
  [arXiv:1412.3094 [hep-ph]].
  %%CITATION = doi:10.1103/PhysRevD.91.055019;%%
  %14 citations counted in INSPIRE as of 21 Mar 2017
  
  %\cite{Chivukula:2014npa}
\bibitem{Chivukula:2014npa} 
  R.~Sekhar Chivukula, P.~Ittisamai and E.~H.~Simmons,
  %``Distinguishing flavor nonuniversal colorons from Z′ bosons at the LHC,''
  Phys.\ Rev.\ D {\bf 91}, no. 5, 055021 (2015)
  doi:10.1103/PhysRevD.91.055021
  [arXiv:1406.2003 [hep-ph]].
  %%CITATION = doi:10.1103/PhysRevD.91.055021;%%
  %9 citations counted in INSPIRE as of 21 Mar 2017
  
  %\cite{Simmons:2015gna}
\bibitem{Simmons:2015gna} 
  E.~H.~Simmons, R.~S.~Chivukula, P.~Ittisamai and N.~Vignaroli,
  %``Separating Dijet Resonances Using the Color Discriminant Variable,''
  arXiv:1507.05868 [hep-ph].
  %%CITATION = ARXIV:1507.05868;%%
  
  %\cite{Chivukula:2015zma}
\bibitem{Chivukula:2015zma} 
  R.~S.~Chivukula, P.~Ittisamai, K.~Mohan and E.~H.~Simmons,
  %``Color discriminant variable and scalar diquarks at the LHC,''
  Phys.\ Rev.\ D {\bf 92}, no. 7, 075020 (2015)
  doi:10.1103/PhysRevD.92.075020
  [arXiv:1507.06676 [hep-ph]].
  %%CITATION = doi:10.1103/PhysRevD.92.075020;%%
  %4 citations counted in INSPIRE as of 21 Mar 2017
  
%\cite{Han:1998sg}
\bibitem{Han:1998sg} 
  T.~Han, J.~D.~Lykken and R.~J.~Zhang,
  %``On Kaluza-Klein states from large extra dimensions,''
  Phys.\ Rev.\ D {\bf 59}, 105006 (1999)
  doi:10.1103/PhysRevD.59.105006
  [hep-ph/9811350].
  %%CITATION = doi:10.1103/PhysRevD.59.105006;%%
  %931 citations counted in INSPIRE as of 31 Jul 2017
  
  %\cite{NWA}
\bibitem{NWA} 
  U.~Haisch and S.~Westhoff,
  %``Massive Color-Octet Bosons: Bounds on Effects in Top-Quark Pair Production,''
  JHEP {\bf 1108}, 088 (2011)
  doi:10.1007/JHEP08(2011)088
  [arXiv:1106.0529 [hep-ph]].
  %%CITATION = doi:10.1007/JHEP08(2011)088;%%
  %74 citations counted in INSPIRE as of 20 Mar 2017

  %\cite{Harris:2011bh}
\bibitem{Harris:2011bh} 
  R.~M.~Harris and K.~Kousouris,
  %``Searches for Dijet Resonances at Hadron Colliders,''
  Int.\ J.\ Mod.\ Phys.\ A {\bf 26}, 5005 (2011)
  doi:10.1142/S0217751X11054905
  [arXiv:1110.5302 [hep-ex]].
  %%CITATION = doi:10.1142/S0217751X11054905;%%
  %70 citations counted in INSPIRE as of 20 Mar 2017

%\cite{Pumplin:2002vw}
\bibitem{Pumplin:2002vw} 
  J.~Pumplin, D.~R.~Stump, J.~Huston, H.~L.~Lai, P.~M.~Nadolsky and W.~K.~Tung,
  %``New generation of parton distributions with uncertainties from global QCD analysis,''
  JHEP {\bf 0207}, 012 (2002)
  doi:10.1088/1126-6708/2002/07/012
  [hep-ph/0201195].
  %%CITATION = doi:10.1088/1126-6708/2002/07/012;%%
  %5191 citations counted in INSPIRE as of 21 Mar 2017

  %\cite{MadGraph}
\bibitem{MadGraph} 
  J.~Alwall, M.~Herquet, F.~Maltoni, O.~Mattelaer and T.~Stelzer,
  %``MadGraph 5 : Going Beyond,''
  JHEP {\bf 1106}, 128 (2011)
  doi:10.1007/JHEP06(2011)128
  [arXiv:1106.0522 [hep-ph]].
  %%CITATION = doi:10.1007/JHEP06(2011)128;%%
  %2444 citations counted in INSPIRE as of 31 Jul 2017

%\cite{Gumus:2006mxa}
\bibitem{Gumus:2006mxa} 
  K.~Gumus, N.~Akchurin, S.~Esen and R.~M.~Harris,
  %``CMS Sensitivity to Dijet Resonances,''
  CMS-NOTE-2006-070, CERN-CMS-NOTE-2006-070.
  %%CITATION = CMS-NOTE-2006-070, CERN-CMS-NOTE-2006-070;%%
  %13 citations counted in INSPIRE as of 21 Mar 2017
    
  
 %\cite{Chatrchyan:2013qha}
\bibitem{Chatrchyan:2013qha} 
  S.~Chatrchyan {\it et al.} [CMS Collaboration],
  %``Search for narrow resonances using the dijet mass spectrum in pp collisions at $\sqrt{s}$=8  TeV,''
  Phys.\ Rev.\ D {\bf 87}, no. 11, 114015 (2013)
  doi:10.1103/PhysRevD.87.114015
  [arXiv:1302.4794 [hep-ex]].
  %%CITATION = doi:10.1103/PhysRevD.87.114015;%%
  %155 citations counted in INSPIRE as of 19 Apr 2017
  
    %\cite{PDG}
\bibitem{PDG} 
  C.~Patrignani {\it et al.} [Particle Data Group],
  %``Review of Particle Physics,''
  Chin.\ Phys.\ C {\bf 40}, no. 10, 100001 (2016).
  doi:10.1088/1674-1137/40/10/100001
  %%CITATION = doi:10.1088/1674-1137/40/10/100001;%%
  %1290 citations counted in INSPIRE as of 02 Aug 2017
  
  %\cite{Das:2016pbk}
\bibitem{Das:2016pbk} 
  G.~Das, C.~Degrande, V.~Hirschi, F.~Maltoni and H.~S.~Shao,
  %``NLO predictions for the production of a spin-two particle at the LHC,''
  Phys.\ Lett.\ B {\bf 770}, 507 (2017)
  doi:10.1016/j.physletb.2017.05.007
  [arXiv:1605.09359 [hep-ph]].
  %%CITATION = doi:10.1016/j.physletb.2017.05.007;%%
  %9 citations counted in INSPIRE as of 21 Aug 2017
\end{thebibliography}
\end{document}